# Miniaturized liquid metal composite circuits with energy harvesting coils for battery-free bioelectronics and optogenetics

Denis Rocha[1], Pedro Lopes[1], Paulo Peixoto[1], Aníbal de Almeida[1], Mahmoud Tavakoli[1]
[1] Soft and Printed Microelectronics Lab, University of Coimbra, Portugal

**Abstract**

Over the past years, rapid progress has been made on soft-matter electronics for wearable and implantable devices, for bioelectronics and optogenetics. Liquid Metal (LM) based electronics were especially popular, due to their long-term durability, when subject to repetitive strain cycles. However, one major limitation has been the need for tethering bioelectronics circuits to external power, or the use of rigid bulky batteries. This has motivated a growing interest in wireless energy transfer, which demands circuit miniaturization. However, miniaturization of LM circuits is challenging due to low LM-substrate adhesion, LM smearing, and challenges on microchip-interfacing. In this article, we address these challenges by high-resolution laser-assisted micropatterning of biphasic LM composites and vapor-assisted LM microchip soldering. Through development of a search algorithm for optimization of the biphasic ink coil performance, we designed and implemented micro coils with trace spacing of 50 μm that can harvest a significant amount of energy (178 mW/cm$^2$) through near field inductive coupling. We show miniaturized soft-matter circuits with integrated SMD chips such as NFC chips, capacitors, and LEDs that are implemented in a few minutes through laser patterning, and vapor-assisted soldering. In the context of optogenetics, where lightweight, miniaturized systems are needed to provide optical stimulation, soft coils stand out in terms of their improved conformability and flexibility. Thus, this article explores the applications of soft coils in wearable and implantable devices, with a specific focus on their use in optogenetics.

**Introduction**

In recent years, the field of wearable and implantable devices has witnessed significant advancements, revolutionizing the monitoring and treatment of individuals[1–4], as well as enabling groundbreaking discoveries in animal studies[5–7]. However, a common challenge faced by these devices is the issue of tethering or reliance on bulky batteries, which often hinders their usability and convenience[8]. This limitation has sparked immense interest in the realm of energy harvesting solutions, motivating research on energy harvesting from light[9–11], mechanical motion through piezoelectric or triboelectric devices[12–14], or far field antennas[15]. Among these approaches, the use of magnetic resonance for energy transfer has garnered significant attention due to its superior power output and the fact that it is not dependent on external light or mechanical pressure[16]. While magnetic resonance coupling is typically performed at short distances, recent studies[17] have demonstrated the potential to extend the range of power transfer to a few meters.

Previous works have focused on flexible circuits with copper-based conductors to produce the coils necessary for energy harvesting. However, the next generation of wearable, and especially implantable devices, demand soft-matter electronics, with stiffness close to that of human organs. If these soft-matter circuits are as well stretchable, they can deal with the strain applied to them due to the dynamic morphology of the human organs, accommodating the mechanical deformations experienced during daily movements[1]. Although copper/Au serpentine-shaped circuits[18] have shown some degree of flexibility, and high electrical conductivity, challenges remain such as fabrication difficulties, the need for clean room lithography, limited stretchability, and lack of intrinsic softness[19,20].

Within the broader context of soft coil applications, this article places special emphasis on a specific scenario: Optogenetics. Optogenetics has risen to prominence as a potent new paradigm for understanding how neurons connect and give rise to more complex brain functions. This enabled rapid scientific progress, leading to the belief that direct clinical use of this technology is forthcoming, potentially leading to improved treatments for a variety of neurological illnesses.[21] Recent applications of optogenetics in neuroscience involve manipulating or monitoring electrophysiological functions and disease models,[22,23] pacing the heart[24] or treatment of bladder dysfunctions.[7]

Conventional means for delivering light to targeted regions in behaving animals uses optical fibers.[24] They are inserted using stereotactic equipment, and external fixtures, surgical glues, cement, and sutures are used to hold them in place.[25–27] However, the high moduli, stiff materials used in these systems are fundamentally mismatched with the soft, compliant tissues of the brain and peripheral nerves.[28] Micromotions brought on by mechanical forces transmitted through the related hardware[29,30] as well as those resulting from natural, biological movements can degrade the biotic/abiotic interface by causing cellular damage and the formation of glial scars.[31] The tethers inevitably affect the animals by restricting their range of motion and altering their behaviours in ways difficult to quantify, which hinders the reproducibility of the investigations and clouds the interpretation of the data obtained.[24] As a replacement for tethered and battery-operated devices, the possibilities of wireless, battery-free and fully implantable devices have recently been explored[7,15]. The result is lightweight neural interfaces with small form factors that enable long-term functionality and better mechanical adaptation to soft neural tissue[28] . However, to provide some degree of stretchability to the devices, recent approaches still rely on deterministic circuit architectures, such as serpentine geometries. The natural body motion of a mouse can lead to compression and elongations up to 60% on implanted devices[32]. Still, the serpentine probe of the head-mounted devices[32] developed recently sustains stretching only up to 30%. Moreover, their fabrication process requires expensive cleanroom lithography with over ten fabrication steps[19].

To wirelessly power an optogenetic stimulator different types of wireless energy harvesting strategies are available such as ultrasound[33], far-field RF radiation (frequency of ~2.4 GHz)[15,34], or near-field power transfer, such as magnetic resonance[35,36]. This article's focus is on magnetic resonance, since with high-quality factor (Q) resonators, efficient power transfer can be accomplished over longer distances (up to a few meters)[16,17]. Moreover, within the frequency range of 100 kHz[37,38] to 200 MHz[39] , the harvested energy is little affected by changes in the environment's dielectric properties, as well as by the existence of obstructions, thanks to the non-radiative nature of the harvesting technique. This is true even when the line of sight between the two resonators is entirely blocked[16]. A more broader use case for wearable wirelessly powered bioelectronics would be its application in the clinical setting for neuromodulation, where one can consider a device that works by magnetic resonance at close range. Thus, the energy requirements differ from the application in long-range optogenetic studies, given that in the latter the magnetic resonance principle has to be extended to allow a larger distance between the receiver and transmitter antennas on the outside of the animals' cages[8,32].

Given the need of interfacing biological tissues in both of the applications mentioned, recent progress has been made in the direction of achieving better conformability and comfort, thus enhanced biocompatibility. In light of this, one can think of implants composed not only by soft polymers in the dielectric encapsulation layers, but also soft electrically conductive polymers as the basis of the electrical circuitry, thus achieving what we term as "fully soft" implants. To accomplish this, it is necessary to develop "soft coils," i.e., coils that are not made of rigid conductor materials such as copper, but of soft and stretchable inks.

Lithography-free approaches have been explored for the rapid prototyping of flexible and stretchable electronics. These include the use of nanowires, which provide stretchability and mechanical durability while avoiding complex fabrication processes[40]. However, nanowires can suffer from long-term stability issues, such as oxidation or fatigue under repeated mechanical stress, which may limit their reliability[41]. Metal nanoparticles are another option, enabling high-conductivity circuits with simple processing techniques, though they may require additional steps for stability and integration[42]. Moreover, metal nanoparticles can exhibit issues like agglomeration, which can degrade their electrical performance over time, and their relatively low mechanical robustness may limit their use in highly deformable applications[43]. Additionally, conducting polymers [44] present a flexible and stretchable alternative, providing ease of fabrication, although their conductivity often falls short compared to metallic conductors. While each of these approaches has its merits, the biphasic liquid metal-based ink used in this work and on refs.[19,20,45] offers an advantageous combination of ease of patterning, high conductivity, and mechanical adaptability, overcoming many of the limitations found in previous methods.

One promising approach is the use of a composite of Eutectic Gallium-Indium (EGaIn) with stretchable polymers, which provides both high electrical conductivity and mechanical deformability[19,20,45]. Renowned for its high electrical conductivity and unique liquid form at room temperature, EGaIn offers several advantages for the realization of soft electronics. However, due to its liquid phase, EGaIn is difficult to deposit, and pattern, and is smearing to touch.[46] Seminal works on laser patterning and laser sintering of liquid metals[47,48], were followed by higher resolution patterning[49] for visually imperceptible electrodes. Still, as the LM is smearing and can easily cause short circuits, achieving a high patterning resolution in ref.[49] required multiple steps of metal deposition and patterning and access to clean room lithography. On the other hand, electrical conductivity limitations are often observed in liquid metal-based materials for antenna applications, prior research has focused on the fabrication of thick metal layers to reduce sheet resistance and thereby enhance overall performance[50–52]. These studies illustrate that increasing the thickness of the metal layer is an effective strategy for overcoming the intrinsic low conductivity of liquid metals. However, increasing the thickness of the metal layer can be a problem for liquid metal-free printed coils due to natural deformations on the body, which can compromise their long-term durability. Furthermore, methods based on liquid metal selective wetting on Cu/Au seed layers[50,53] have enabled high resolution liquid metal based stretchable circuits. However, there is a limitation on the achievable circuit thickness, as the thickness of the deposited liquid metal depends on the width of the seed trace, thus reducing the control over the applied liquid metal thickness. The trace thickness and resolution are as well limited due to the high probability of short circuits between the adjacent liquid metal traces, which is further problematic during microchip interfacing. Therefore, liquid metal wetting techniques are not yet able to simultaneously achieve high resolution patterning, high trace thickness, and reliable microchip interfacing. On the other hand, the techniques developed in this work not only address these problems, they as well eliminates the need for clean room deposition and lithography, and manual deposition processes, such as liquid metal wetting or HCl vapour for SMDs interfacing, making it lower-cost, more scalable and more reproducible compared to previous works.

We previously developed versions of digitally printable biphasic composites based on liquid metal and conductive silver composites which have been extensively discussed in ref.[19,20,45] The digital printing technique enables rapid fabrication of biphasic ink composites, offering a cost-effective and customizable approach to circuit production. Despite these advancements, there are still challenges that need to be addressed to achieve a fully integrated and battery-free circuit with optimized performance. The resolution of current digital printing techniques may not be sufficient to develop compact circuits required for miniaturized applications. Advancements in rapid laser patterning of liquid metal composites with metal nanowires have improved mechanical stability and electrical conductivity. Recent research[40,54] has also demonstrated enhanced electrical performance and flexibility through the integration of these materials. Other studies [55] reveal high-resolution laser-patterning techniques that maintain mechanical integrity. Moreover, laser patterning stands out[53,56,57], as a technique for high-resolution patterning. However, implementation of wireless optogenetic devices demands for a combination of high-resolution patterning, microchip integration, and high conductivity traces, which remains a challenge.

In this work, by combining the use of biphasic liquid metal-based composites, high-resolution laser patterning, and the vapor-soldering technique[19], we demonstrate miniaturized hybrid circuits based on liquid metal, that can be used for optogenetics applications. This research focuses on rapid prototyping of soft-matter coils for energy harvesting through magnetic resonance coupling. To afford the desired low stiffness and stretchability for the circuits, we use a highly stretchable and digitally printable biphasic liquid metal-based conductive composite[20,58]. Liquid metals, including Eutectic Gallium Indium (EGaIn) have proven to provide the best trade-off between electrical conductivity and mechanical deformability. Using digital fabrication techniques such as laser patterning, we demonstrate simultaneous fabrication of coils along with other parts of circuits in minutes and with accessible equipment, resulting in cost-effective and rapid implementation of customized circuits that can be adapted according to the user and application[19,20]. This work builds over these previous works, by patterning a biphasic ink that, despite containing liquid metal, has a non-smearing behaviour and therefore its deposition and patterning is straightforward. Moreover, it is as well demonstrated interfacing miniaturized surface mount silicon chips, that are necessary for functional energy-autonomous implants capable of harvesting power from their surroundings through wireless power transfer[38]. We also investigate the optimal geometry to maximize energy harvesting, given the maximum patterning resolution, conductivity of the biphasic ink and minimizing its overall size due to the space restrictions associated with implantable devices.

This article outlines the methodology for fabricating these soft coils, evaluates their performance, and discusses the challenges and future directions for integrating fully autonomous, wireless energy-harvesting systems in wearable and implantable bioelectronic devices.

## Results

***Fabrication of soft circuits with integrated energy harvesting coil and microchip assembly.*** In order to fabricate the battery-free soft circuit with integrated microchip, three main steps should be performed: deposition and patterning of the soft conductive circuit, integration of the SMD components, and encapsulation of the circuit with a biocompatible polymer. Figure 2A summarizes this process. Here we use a biphasic EGaIn-Ag composite previously developed by our team[20] as the conductive ink. As patterning options, various methods exist such as extrusion printing, laser patterning, or stencil printing. Previously we demonstrated that the ink can be digitally printed through extrusion printing[20]. While this permits rapid additive manufacturing of circuits, the printing resolution of ~300μm (trace width and spacing) is limited for miniaturized coils[7,15]. Therefore, we used a laser patterning technique in order to obtain more compact coils. This permitted improving the resolution significantly to trace widths of 150 μm and spacing between turns of 50 μm, and consequently fitting more coil turns in the limited space.

Two similar biphasic EGaIn-Ag composite inks were used. These composite inks were studied in the articles[20,51], they have only a difference in the polymer used as substrate. The AgEgaIn-TPU (up to ~$1.08 \times 10^6$ S/m), was also employed in this study due to its slightly higher electrical conductivity than AgEgaIn-SIS (up to ~$4.56 \times 10^5$ S/m), which may be attributed to a more efficient evaporation of the TPU solvent, THF:DMF when compared the SIS's toluene[59,60]. This results in better packing of the composite, thus improved percolation. However, the SIS based ink has an already proven good reliable performance on the gluing of the chip-pad interface[19], which is expected since the SIS hyperelastic binder, has excellent adhesive properties. Therefore, as we showed in previous works, when the circuits are subjected to the solvent-assisted soldering process, the substrate fuses with ink, which in turn glues the chip's pad to the circuit, all due to the SIS phase transition.

The process starts with the application of a thin sacrificial film of poly(vinyl alcohol) (PVA) (~300 μm) on a flat surface, such as a glass plate, using a thin film applicator (supplementary fig. 1A). This layer serves as the sacrificial layer that permits easy circuit release. On top of it, layers of ink are deposited using the same thin-film applicator or a metallic roller with tape spacers (supplementary fig. 1B). We used a Master Oscillator Power Amplifier (MOPA) fiber laser with Infrared (1064 nm) wavelength for patterning the circuit (supplementary fig. 1C). This is a low-cost laser system commonly used in print shops to engrave metals. As the laser wavelength mostly affects metals, the PVA layer under the ablation area remains intact. After patterning the circuit, a solution of SIS hyperplastic binder is poured over the patterned circuit, then, after curing and the circuit is peeled off (supplementary fig. 1D). After dissolving the PVA layer, the circuit is turned upside down and SMD components are placed using a manual or automatic pick & place machine (supplementary fig. 1E).

The soldering is performed using a vapor-assisted soldering technique recently developed by our team[19]. A polymer-gel transition is induced on the ink and on the SIS substrate by exposing the circuits to a solvent vapor chamber for 45 min (supplementary fig. 1F). During the gel state, the conductive pads of the microchip package penetrate and adhere to the conductive ink. The bottom side of the package also adheres to the underlying substrate. And since the chip penetrates the substrate, it is completely surrounded on all four sides by the adhesive polymer. Providing at the same time self-soldering and self-encapsulation of the electronics. Finally, the circuit is encapsulated by PDMS, with the help of a 3D printed mold (supplementary fig. 1G).

**Coil design optimization and tuning.** First showcased by Nikola Tesla[61], near-field power transfer utilizes non-radiative electromagnetic energy within approximately λ/2π distance from the transmitter. This method relies on inductive coupling, establishing a connection between a transmitting coil and a receiving coil[62]. This strategy provides a highly efficient mean of transmitting power over small distances, making it feasible for practical implementations in commercial contexts such as wireless charging of electric vehicles[63] and powering of consumer electronics[64].

Although designed for near range, the efficiency of power transfer and the achievable distance can be substantially improved by utilizing transmitting and receiving coils (resonators) that are finely tuned to establish a robust magnetic resonant coupling[16,17]. Thus, coil design optimization and tuning to resonate in the same frequency of the transmitter are crucial (13.56 MHz). In this section, a description of the methodologies followed for the optimization of the coil design, and its subsequent tuning, will be laid out.

*Design optimization.* In the context of magnetic resonance, the coil can be regarded as an inductor[65,66], thus we can evaluate its power harvesting capabilities in terms of its quality factor, $Q = 2\pi f L/R$, being f the frequency of operation, L, the coil's inductance and, R, the coil's resistance. The Q-factor of an inductor is a measure of how close the inductor is to an ideal inductor, i.e. one that has no parasitic resistance. As can be seen in fig. 3A, the receiver coil in the fabricated devices can be defined by 5 parameters: overall width, overall length, number of turns, trace width and trace spacing. These parameters and their relationships define the inductance and resistance of the coil. To optimize them, calculations were performed using the formulas from ref.[67]

The best value for a coil's overall width and length can arguably be the one easier to choose. For maximum power transfer efficiency, the coil should be as big as possible (without surpassing the transmitting coil's size). For wearable applications and for implants, on the other hand, we desired them to be as small as possible, to have the minimum possible impact on the body. So, given the size constraint and considering the fabrication resolution limitations, one can optimize design parameters of the coil such as number of turns, trace width and trace spacing.

An exhaustive search algorithm was developed to thoroughly investigate all potential combinations of values for the number of turns, trace width and trace spacing of the coil. The algorithm systematically assessed the performance of the coil (which is reflected in its Q-factor) by rigorously evaluating each possible combination of values in a defined interval of values for trace width and number of turns. The formulas implemented in the code for the inductance and resistance calculations are present in supplementary formulas 3 and 4. The results were displayed in the form of 2D plots (inductance/resistance/Q-factor versus trace width and number of coils) and intensity maps of the best combination of design values. Subsequently, the obtained data was analysed and visualized to gain insights into the coil's performance across a range of value combinations. This methodology enabled a meticulous examination of the parameter space, aiding in the selection of the most suitable design for enhanced performance and efficiency given the fabrication limits in resolution (which were about 50μm for trace width and spacing).

In the search for the optimized values, we imposed conditions such as the exclusion of parameters combinations that resulted in: Q values below 25; resistance values above 3 ohms; inductance values below 0.7 μH and inner diameters smaller than 2 mm. In the end, all these criteria were combined and evaluated in a single plot (fig.3Bi). To visualize the results of the search algorithm, intensity maps and 2D plots were created (figs.3B). As one can see in fig.3Bi there is a solution that maximizes the Q-value, but it can't be taken alone as the only evaluation parameter. Moreover, around the maximum predicted Q-value there is a parameter space where the differences in this value are marginal. Thus, a more holistic approach should be taken.

Starting by the inductance value (fig.3Bii), one can see that it increases with the increase of the number of turns since it can concentrate more of the electromagnetic field. On the other hand, maintaining the number of turns fixed, the inductance decreases with the increase of trace width, which can be explained by the fact that the turns get more on the outer contour of the coil, which in turn decreases the coil's fill ratio[65–67]. This insight is of major importance when designing small coils, which due to their space constraint have difficulty in achieving high inductance values, thus reducing the trace width helps to improve the Q-value of a coil. However, the trace width can only be reduced to some extent until it stops being beneficial[65,67], because of the fact the resistance also increases with its reduction (fig.3Biii). On the other hand, miniaturized coils have the benefit of much lower trace lengths for the same number of turns when compared to coils of bigger sizes.

Finally, the next parameter to evaluate is the coil's inner diameter, i.e. the empty space inside the coil (measured on its last inner turn). This parameter is also associated with the coil fill ratio[65–67]. This parameter gains substantial importance when designing miniaturized coils since often it will be the space where the electronics will stay, so depending on the circuit complexity this space can be adjusted. In fig.3Biv, one can see that, as expected, the increase in the number of turns or the trace width will always result in an increase in the inner diameter. However, its minimization is not detrimental to the coil's Q-value until inner diameters of around 1 - 2 mm[65–67].

*LC resonator tuning.* Once the optimum parameters for manufacturing the coils had been found, it was necessary to find the right tuning for them. Coil tuning for magnetic resonance refers to the process of matching the electrical properties of the coil to the desired resonant frequency [16,17]. It involves optimizing the coil's inductance (L) and capacitance (C) values to create a resonant LC circuit. Tuning is necessary to maximize the efficiency of the coil. It can be made simply by adding capacitors in parallel between the coil and the load[16,17].

The exact values of the required capacitance were determined using a vector network analyzer (VNA) by two different methods: direct measurements, where the coil was directly connected to the VNA (figs.3E) and indirect measurements, where the coil was not connected to the VNA, but was placed on top of a copper coil with similar dimensions, being this one connected to the VNA (fig.3D).

With the direct measurements, we were able to get the S11 response of the system coil and tuning capacitors for various values of the capacitors until the curve's dip was aligned with 13.56 MHz. Then the LED (necessary for optical stimulation in optogenetic application) was also added in parallel. Since it can be seen as a capacitor itself by the coil, its capacitance is also important to ensure the finding of the correct values for the capacitors to be added.

With the direct measurements we obtained the curve of inductance as a function of frequency (Fig.3Ei) and the curve of impedance moduli as a function of frequency (Fig.3Eii). By experimenting with different values for the tuning capacitors, the optimum tuning can be obtained, i.e. when the peak of the curve is at 13.56 MHz, in the case of the curve of the impedance moduli versus frequency. For the case of the inductance versus frequency curve, it is when the point of zero inductance is aligned with 13.56 MHz, i.e., the point that minimizes the inductive reactance[65,66]. This method was mainly used for the smaller designs, as their small factor means that they cannot cause any measurable changes in the magnetic field of the receiver coil, which is connected to the VNA in the direct measurement method.

*Performance comparison of copper versus soft coils.* For the evaluation of the performance of copper versus soft coil, four coils were selected from supplementary table 2. The coils selected were the ID1, ID2, ID3, and ID4 (fig.4A). ID1 is a commercial coil from Molex (model 1462360051); ID2, fabricated in fPCB, has 9 turns, 150 μm trace width, 50 μm trace spacing and overall size of 10 mm diameter; ID3, fabricated in soft materials, has 9 turns, 150 μm trace width, 50 μm trace spacing and overall size of 10 mm diameter (same design as ID2) and ID4, fabricated in soft materials, has 9 turns, 150 μm trace width, 50 μm trace spacing and overall size of 13 x 7.6 mm.

The antennas ID1 and ID2 were selected as high-performance counterparts for comparison with the best soft antennas fabricated in this work, the ID3 and ID4. It's worth noting that the conductivity of printed inks is ~7.02 × 10$^5$ Sm$^{-1}$ which is

<80× compared to the copper. Therefore, achieving the same efficiency of commercial coils with printed coils is very difficult. Nevertheless, printed coils are attractive due to fabrication advantages, and the possibility of development of soft, and ultrathin electronics that are necessary for wearable and implantable devices.

A metric was created to compare the performance of the coils, named the figure-of-merit of miniaturization, $FOM_{miniaturization}$:

$$FOM_{miniaturization} = \frac{L \cdot V_{pp}}{A \cdot \rho} \cdot 100 - R$$

Being L the inductance (µH) and Vpp the peak-to-peak voltage (V). The A represents the area (mm$^2$) and ρ is the coil fill ratio. The resistance represented by R is expressed in ohms.

$FOM_{miniaturization}$ serves as a comparator metric which was designed for finding the best coil design when the physical space (for implantation, in this case) is restricted. Three main variables are involved: coil dimension, coil inductance, and coil resistance. In an ideal world, one would strive for the largest possible coil given the space limitation, with the highest possible inductance and lowest possible electrical resistance. However, for example, when one tries to improve the Q-factor by increasing the inductance through increasing the number of turns or decreasing the trace width, an inevitable increase in the electrical resistance will happen, which has a negative effect on the Q-factor, hindering the desired gains. The same applies for when one tries to increase the Q-factor by increasing the overall dimensions of the coil. The size limitation for a coil implanted on the mice's head (~10x12mm, ref.[32]) is higher than on its back (where implants exceeding 200 mm2 have been demonstrated[32]), thus, a head-mounted coil will be a lower burden for the mice than a back-implanted one. Hence, a metric was needed for comparison between coils of different sizes. Therefore, the objective of the $FOM_{miniaturization}$ is to highlight which coil designs provide a more efficient power harvesting performance with the smallest possible footprint. The $FOM_{miniaturization}$ formula was created so that it values high values for inductance and a peak-to-peak voltage in the same proportion that awards low areas and coil fill ratios, but not overly emphasizes on the resistance. Given that when looking to fabricate liquid metal stretchable/deformable coils, a compromise has already been established between electrical conductivity and device deformability.

To obtain the data of the plots in fig. 4E-G, three setups were used: (fig.4B,4C, 4D). Setup 1 consisted of connecting a half-bridge circuit connected to the coil to use it as a rectifier to test the response of the coils to different loads. In this setup, the receiver coils were in close contact with the transmitter coils. Setup 2 had the receiver coils directly connected to the oscilloscope to evaluate the variation of the peak-peak voltage of the AC wave induced by the transmitter coil. Here the distance between the transmitter and receiver coil was changed with the help of an F-shaped clamp clip and this change in distance was measured with a laser meter. Setup 3 was broadly similar to experimental setup 2, with the exception that a piece of tissue from a C57/BL6 mouse was inserted between the transmitter and receiver. This is to simulate a case in which the receiver is implanted, and the transmitter is placed over the skin as a wearable textile or printed circuit (fig. 4D). Therefore, the implant coil was placed in contact with the tissue.

If we analyze the graphs, we can first see from Fig. 4E that the ID1 and ID2 antennas were able to achieve a maximum harvest power of more than 180 mW. In these measurements, the coils were in close contact with the transmitter coil. Coil ID3 has the same design as ID2, so these two coils are useful for analyzing the difference in power harvesting between soft antennas and copper-based antennas. The resistance of coil ID3, which was made with conductive ink, was ~21.80 Ω, while the resistance of its copper counterpart, ID2, was ~2.44 Ω. So, although the resistance of the coil was increased by ~9 times, ID3 was still able to harvest more than 140 mW at its peak. Also, this is ~25% less than the copper coil, it is still a very good value, considering the fact that the coil is printed using an ink. Various optimizations were performed to achieve a coil with this value of electrical resistance. These iterations are listed in the supplementary table 3-5. The improvements that were added to the manufacturing method are also listed there. The coil ID4 had a different design from the one of antennas ID2 and ID3, however, it had an inductance (1.37 µH) close to what the antennas ID2 and ID3 had achieved. But, since its coil length was also bigger (294.34 mm, versus 227.81 mm of ID3 coil), it had a bigger resistance (46.55 Ω). This resulted in a peak of harvested power of around 100 mW, substantially lower than the other three.

In terms of the effects of distance between the transmitter and receiver coil, as one can see in the plot of figure 4F, antennas ID1-3 started at the same voltage level (~21 V), whereas ID4 already started at a lower voltage (~13 V). All the coils' harvested voltage values seem to have an exponential decay with the distance. After 1.5 cm, none of them demonstrated a peak-to-peak voltage above 1.5 V. Which was expected due to the size of the transmitting coil, which was comparable to the ones of the receiving coils.

Regarding how the fabricated coils are affected by the presence of tissue in close proximity (fig.4G), i.e. the capacity to simulate the coil being implanted under the skin and the transmitter coil as wearable on top of it, by analyzing the plot of fig. 4G is perceptible that between 0.2 − 1.2 cm of distance from the transmitter coil, the differences in power harvesting with and without tissue seem to be negligible. Overall, the tissue didn't seem to have much effect on the coil's energy harvesting capabilities. The biggest difference occurred when the tissue was in close contact with both the receiver coil and the transmitter coil. In this case, the receiver coil is only capable of harvesting about 14 V, while the measured voltage without the tissue was ~18 V. This difference is partly due to the fact that the thickness of the tissue did not allow the coils to be as close together as is possible when it is not present. Nonetheless, due to the non-radiative nature of magnetic resonant coupling, it was already expected that this type of wireless power transfer was going to be relatively unaffected by changes in the environment's dielectric properties as well as by the existence of obstructions[16,37,38].

To demonstrate the capabilities of liquid metal composite-based coils for wireless battery-free optogenetics, we performed long-range experiments in a 35 x 25 cm cage with a double loop copper coil wrapped around for wireless power transmission (fig. 5E-G). Here we added the concept of multi-agent optogenetic control (illustrated in figure 1), consisting of wireless energy transfer in a cage of mice, where all of the optogenetic implantables can be controlled on-the-fly. That is, multiple battery-free passive coils can be simultaneously controlled through a single wireless "reader" device. Data regarding the energy harness by the coils in different locations of the cage are available in figure 5G.

## Discussion

In this work, we demonstrated materials and methods for rapid prototyping of soft-matter implantable coils for energy harvesting for optogenetic neuromodulation. We used a liquid metal-based ink to achieve a high electrical conductivity, and a laser patterning technique to achieve rapid and low-cost prototyping. Through several evolutions and optimizations of the fabrication method (supplementary tables 3-5), it became possible to fabricate soft coils with 150 μm trace widths and 50 μm trace spacings. This permitted the development of smaller circuits. The overall size of the first device produced was reduced to ~ 1/7 of the first coil produced by digital printing, from ~600 mm$^2$ (first prototype) to 78.5 mm$^2$ (final prototype), resulting in a final device smaller than 1 cm$^2$.

The improvements achieved in this work on the fabrication method are also of great value for the adaptation of the designed devices to other applications such as photometry for monitoring neuronal dynamics in the deep brain[6,68] and local tissue oximetry for continuous sensing of local haemoglobin dynamics[5]. Other clinical applications in the area of neuromodulation for other organs can be as well envisioned, for example, for the application of optogenetics for cardiovascular medicine[28,69].

***Towards wireless optogenetics for neuroscience research.*** The miniaturized implant of fig. 5E can be used for wireless optogenetic stimulation in behavioral experiments. Its footprint (19 x 11 mm) has dimensions comparable to the ones used in ref.[32] to implant copper-based rigid devices in small mice breeds such as C57BL/6, which are widely used in animal research.

However, the implants developed by ref.[32] are fabricated on the basis of rigid conductors such as copper and flexible substrates such as polyimide (Dupont Kapton®, which has a Young´s Modulus of ~2.5 GPa (ref.[70])), whereas the skin's Young´s Modulus is 1-10 kPa (ref.[71]). The use of Cu-based circuits results in devices that lack tolerance to strain and as well lack mechanical compliance with the host, making it uncomfortable, thus failing to adapt to the animal body (e.g. mice C57BL/6 or Balb/c, breeds used predominantly in optogenetics research) much less to their large range of movements. This will inevitably cause stress in the animals[24,28], which in turn will invalidate the conclusions obtained from the animal experiments conducted. On the other hand, the implants that we developed tackle this challenge through enhanced mechanical performance due to their stretchability/deformability, without resorting to lithography techniques which are expensive and require the use of cleanrooms.

To our knowledge, this is the first demonstration of miniaturized liquid metal-based coils being wirelessly powered in a long-range setting, in a cage with dimensions comparable to those required to perform wireless optogenetics experiments in freely moving mice. Finally, further development will include a probe system with micro-sized LEDs to transmit electricity and light into the brain, as in ref.[32].

***Towards optogenetic sacral neuromodulation.*** The miniaturized implant in fig. 2C can be developed as an optogenetic stimulator for sacral neuromodulation in overactive bladder (OAB) patients. This requires a probe, as OAB stimulators are typically subdermal and use probes to conduct electricity to inner nerves, making the surgery minimally invasive[72]. Alternatively, placing the stimulator near the nerves complicates communication and increases invasiveness. For optogenetic stimulation, the probe can either conduct electricity to a μLED or use an optical fiber to conduct light. The implant can be wirelessly powered and communicate via inductive coupling, requiring the patient to wear a powering patch, or it can have a rechargeable battery charged by the patch. This work lays the foundation for developing fully soft optogenetic implants for neuromodulation in diseases like OAB, heart arrhythmia, Parkinson's, and Alzheimer's.

While the developed devices are promising for optogenetic applications, it is important to address the biocompatibility of the materials used in the implants. In this case, the liquid metal (Eutectic Gallium-Indium, EGaIn) is fully encapsulated in PDMS, rather than being directly embedded in the tissue. Therefore, the primary material to study regarding biocompatibility should be PDMS, which has been extensively researched and shown to have favorable biocompatibility properties for use in biomedical applications[73]. For instance, studies have demonstrated that PDMS exhibits good compatibility with various cell types and is well-tolerated in vivo. However, the potential cytotoxic effects of the liquid metal itself, particularly with prolonged exposure, should not be overlooked as they may pose challenges in the long-term functionality of implantable devices[74,75].

## Methods

**PVA synthesis:** The PVA solution is obtained by mixing 5g of PVA powder (Selvol 125, SEKISU) with 50 mL of $H_2O$. The mixture was stirred at 90 °C using a hot plate until it became a clear and homogeneous solution.

**SIS-based inks fabrication:** The first ink experimented with was a bi-phasic ternary Ag-In-Ga ink developed by ref.20 The AgEGaIn-SIS ink is synthesized by mixing EGaIn, Ag micro-flakes, and styrene isoprene block copolymers (SIS). The ink is prepared by first dissolving SIS block copolymers (15 wt % SIS) in a toluene solution and mixing it in a planetary mixer (30 min at 500 rpm). For each 5g of BCP solution, 6.2g of Ag flakes (Silflake 071 Technic Inc.) were added. However, the proportion of EGaIn:Ag was varied, and tests using ratios of 1, 2, and 4 were made in an attempt to increase the ink's electrical conductivity. The EGaIn solution was prepared by mixing 75.5 wt% Gallium and 24.5 wt% Indium and letting it naturally dissolve for ~12 hours. After the addition of Ag and EGaIn, the solution goes again to the planetary mixer for 3 min at 2000 rpm.

**TPU-based inks fabrication:** The AgEGaIn-TPU ink was synthesized by mixing EGaIn, Ag micro-flakes, and thermoplastic polyurethane (TPU), instead of SIS. The first step of the ink synthesis is dissolving the TPU filament (Ninjaflex TPU 1.75mm 3DNF0817510). A mix of tetrahydrofuran (THF, Sigma Aldrich) with dimethylformamide (DMF, Sigma Aldrich) is used as solvent, and the ratio between 2-mTHF and n-DMF is 4:1. The dissolution of TPU filament (15 wt % TPU) in THF:DMF takes roughly about 4 days at ambient conditions if one just let it dissolve naturally. For each 1 g of TPU solution, one must add 1.75 g of Ag micro-flakes and 1.75 g of EGaIn. Initially, every time a component was added, the solution would go to the mixer (3 min at 2000 rpm). However, when the solution only went to the mixer one time (after all components were added), the final ink seemed to be less viscous, thus easier to spread. Hence, this method was used primarily. Also, the ratio of EGaIn to Ag micro-flakes was 1:1 in the first ink produced, however, in order to increase the ink's conductivity different EGaIn:Ag ratios were experimented (more specifically 2:1 and 3:1).

**SIS substrate preparation:** The SIS solution is prepared by dissolving SIS (Styrene 14%, Sigma Aldrich) in toluene with a ratio of toluene:SIS of 2:1 and mixing it for 30 min at 500 rpm.

**PDMS synthesis:** The PDMS used was Sylgard 184 (Dow Corning Corporation). For 10 g of PDMS used, 1 g of curing agent is required. The solution is mixed for 3 min at 2000 rpm and then degassed for 4 min at 2200 rpm.

**Toluene vapor exposure:** The circuit is placed in an enclosed chamber that contains a paper soaked with toluene, thus generating toluene vapor there. Different treatment times were experimented (more information on supplementary table 3-5).

**PVA and ink deposition:** The PVA layer is deposited with a thin film applicator and it is cure in the oven at 60 °C for about 30 min. Afterwards, two sheets of metallic tape must be placed on the sides of the glass (supplementary fig. 3.2b). They serve as spacers to then apply 2 or 3 layers of conductive ink with a spatula. Between layers, one must let the ink dry in the oven for a few minutes.

**Laser patterning:** A thin film applicator is used to apply a thin layer of conductive ink to the substrate. A pulsed fiber laser (1064 nm wavelength) then selectively removes the ink to separate the circuit traces. The polymer substrate beneath the ablation area is unharmed since this laser wavelength affects mostly metals. This proves to be a successful technique for creating high-resolution circuits with trace spacings as small as 50μm. The parameters used for ablation involved hatchings with a line width of 0.01 mm both in 0º and 90º directions, with the marking of the contour. The power of the laser was set to 100%, its speed to 1000 mm/s with a loop count of 10. The frequency used was 50 kHz with a Q Pulse Width of 40 ns.

**Flexible PCB fabrication:** The process started by applying a layer of photosensitive paint (POSITIV 20) to a fPCB. Then, the next step was laser patterning of the circuit (pulsed fiber laser with 1064 nm wavelength), uncovering the photoresist where one needs the copper to be etched. Next, copper etching of the exposed parts using ferric chloride in a hot plate (AGIMATICN) for 45 min at 40°C. Finally, to strip the rest of the photoresist the fPCB was washed with acetone.

**Electronic device components:** The RF harvesting module was built with a tuner capacitor and the dynamic NFC-accessible EEPROM (1.8 x 2.6 mm, NTAG5, NTP5210, NXP) which provides internal AC-DC rectification, as well as, user-selectable fixed output voltages. Its power consumption in NFC passive communication is 0.66 mW. An older version of the NTP5210 IC, the NFC Forum Type 2 NTAG I²C 1K (NT3H1101W0FHK. NXP), was also used. For the related circuitry components, such as decoupling capacitors, LEDs, and resistors, the 0402 package was selected to minimize the overall footprint of the devices. An SMD red LED (150040RS73240, Wurth Elektronik) was used as the light source in the various device designs, due to its low forward voltage of 2V.

**Wireless RF power transmitter:** The model CLRC663 plus NFC Frontend Development Kit (OM26630FDK) from NXP was used. It is a multi-protocol NFC frontend with a maximum output voltage of 1.9W.

**Coil optimization calculations:** These calculations and plots were obtained with the aid of MATLAB software. Formulas used are present in supplementary materials (supplementary formulas 3 and 4).

**VIA stretchability evaluation:** For electromechanical testing, we used an Instron 5969 with a 100 N load cell and a data acquisition system composed of a and 16-bit DAQ (NI USB 6002) and a multimeter (gwInstek gdm-8351).

**Tissues for the performance comparison tests:** The tissue was collected from the mice (Charles River) used for a biocompatibility study realized by CNC (Centro de Neurociências e Biologia Celular) from the University of Coimbra. All experiments were carried out with the approval of their animal ethics committee (Orgão Responsável pelo Bem-Estar dos Animais (ORBEA)), the approval of the Direcão-Geral de Alimentacão e Veterinária (DGAV), and in accordance with EU directives regarding animal use in research. The tissue was obtained from 2 - 5 months old C57/BL6 mice, had a thickness of ∼ 0.1 mm and a size of ∼ 16 x 18 mm.

**Abbreviations**

Block co-polymers (BCP)
Flexible printed circuit board (fPCB)
SIS-based polymeric Ag and EGaIn conductive inks  (AgEGaIn-SIS)
TPU-based polymeric Ag and EGaIn conductive inks  (AgEGaIn-TPU)
Liquid Metal (LM)
Styrene-isoprene-styrene (SIS)
Light-Emitting Diode (LED)
Eutectic Gallium-Indium (EGaIn)
Vector Network Analyzer (VNA)
Overactive bladder (OAB)

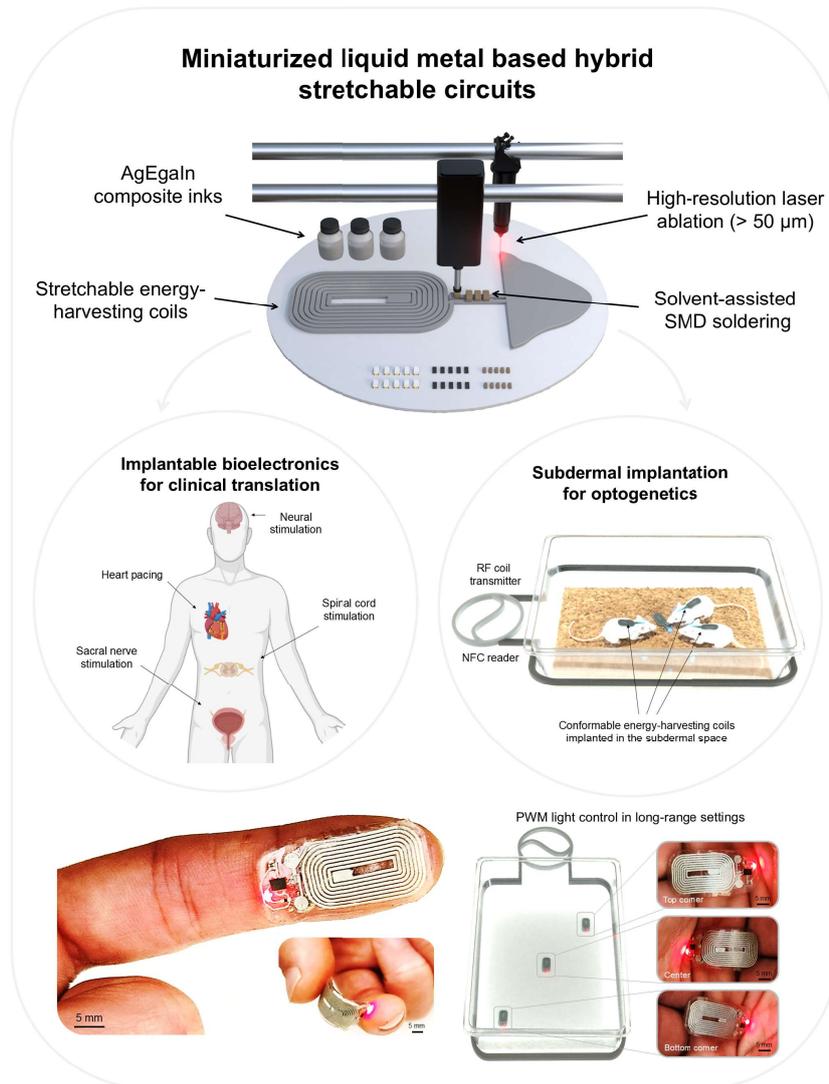

**Figure 1.** Potential applications for wirelessly powered soft electronic implants as stimulators for future translation to the clinical setting and behavioral neuroscience research.

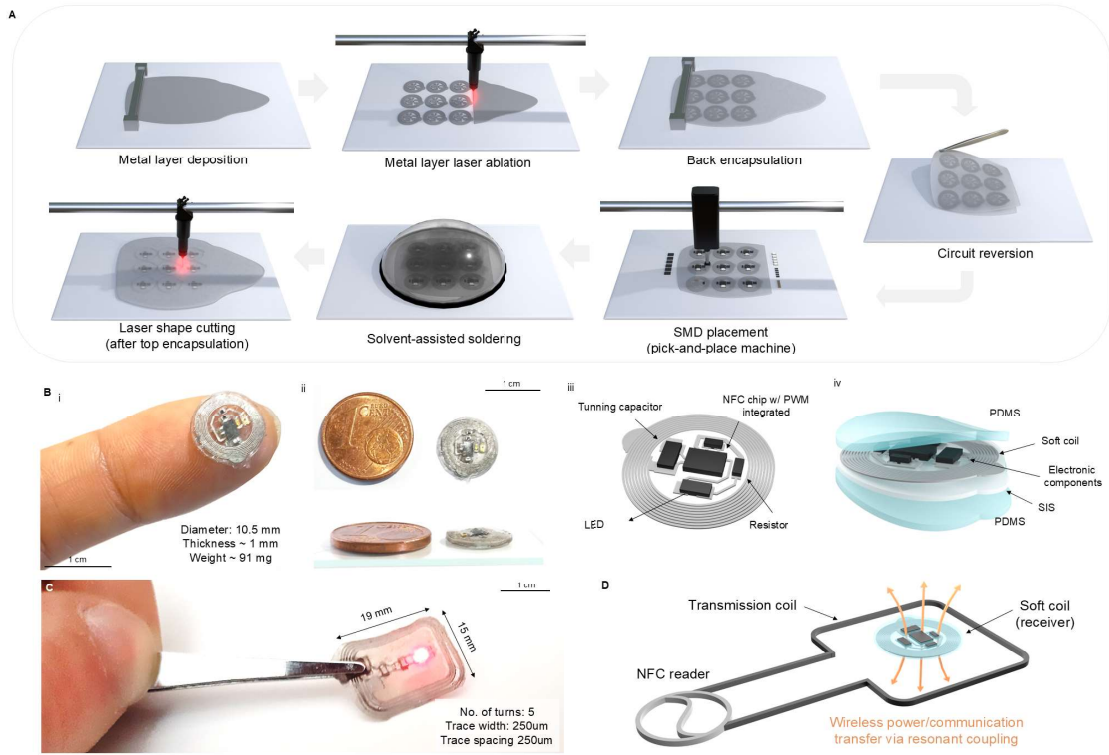

**Figure 2.** Fabrication process and devices developed for near- and long-range applications. (A) Schematic renderings of the main fabrication steps. (B) Miniaturized device developed for near-range applications. Photographs comparing its size to a finger (B) and to a 1 Euro Cent (ii); (iii) rendering of the device's circuit without the encapsulation layers; (iv) rendering of the device encapsulated. (C) Photograph of the miniaturized device developed for long-range applications receiving energy via wireless power transfer. (D) Schematic rendering illustrating how the wireless power transfer/communication transfer works.

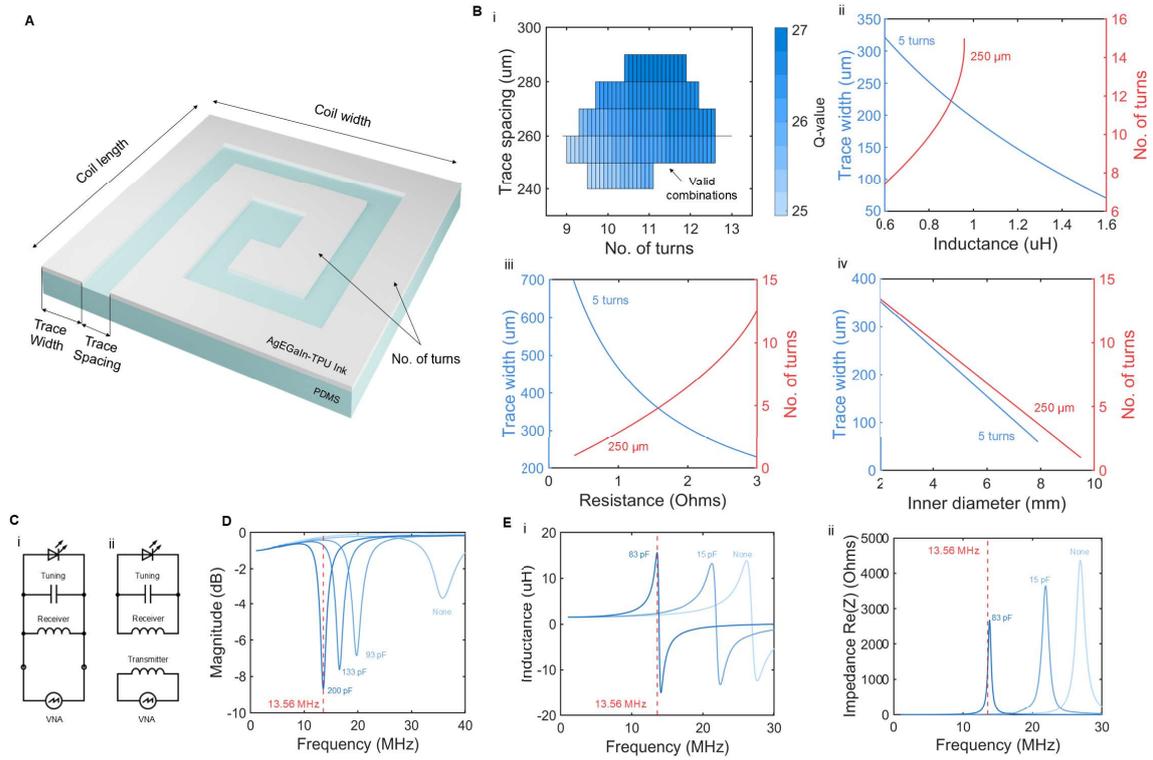

**Figure 3.** Coil's design and tuning optimization. (A) Rendering of a soft coil illustrating the design parameters needed to be considered for its optimization. (B) Intensity map are calculated to find the best combination of values for the number of turns and trace width that optimize the coil's Q-value in (i); plots of the variation of inductance (ii), resistance (iii) and inner diameter (iv) according to trace width (for a fixed value of 5 turns) and number of turns (for a fixed value of 250 um of trace width); (C) Schematics illustrating the setups for the (i) indirect and (ii) direct measurements used for finding the fabricated coils' tuning capacitors values. (D) VNA indirect measurements for a soft coil (design parameters: 19x15mm, 5 turns, 250 um as trace width, and 250 um as trace spacing). (E) VNA direct measurements of a soft coil (design parameters: 10x10mm, 9 turns, 150 um as trace width, and 50 um as trace spacing). In (i), it's displayed the inductance versus frequency curve and in (ii), it is displayed the impedance moduli versus frequency curve. Note: To check the differences of tuning due to the NFC chip used (NTP5210, equivalent to 15 pF, according to the manufacturer), it was also connected in parallel for testing.

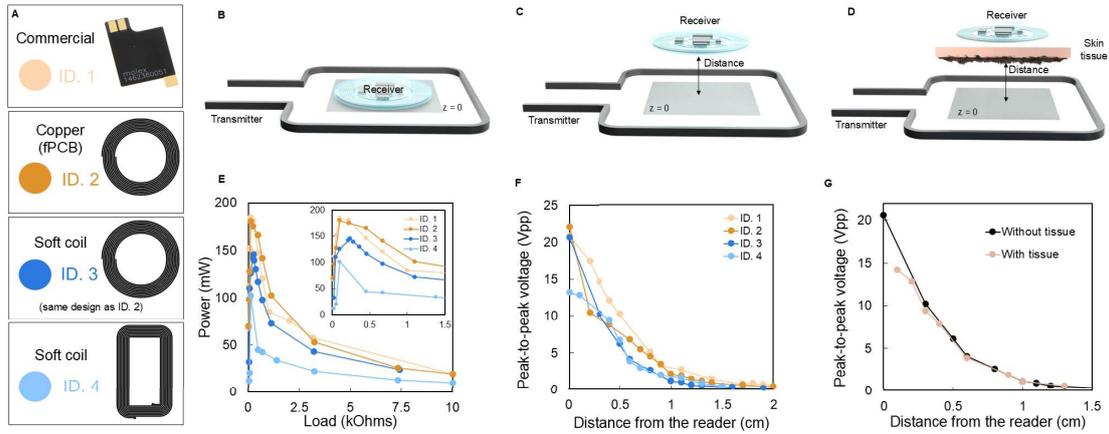

**Figure 4.** Performance comparison between copper and soft coils. (A) Coil designs selected for the study (more information on supplementary table 2). (B) Coil setup used to obtain data for the power versus load plot on (E). (C) Coil setup used to obtain data for the plot on (F) which analyzes the peak-to-peak voltage variation due to the coils' distance from the reader. (D) Coil setup used to obtain data for the plot on (G), where it was analyzed of the effect of the mouse tissue on the peak-to-peak voltage regarding the distance from the reader.

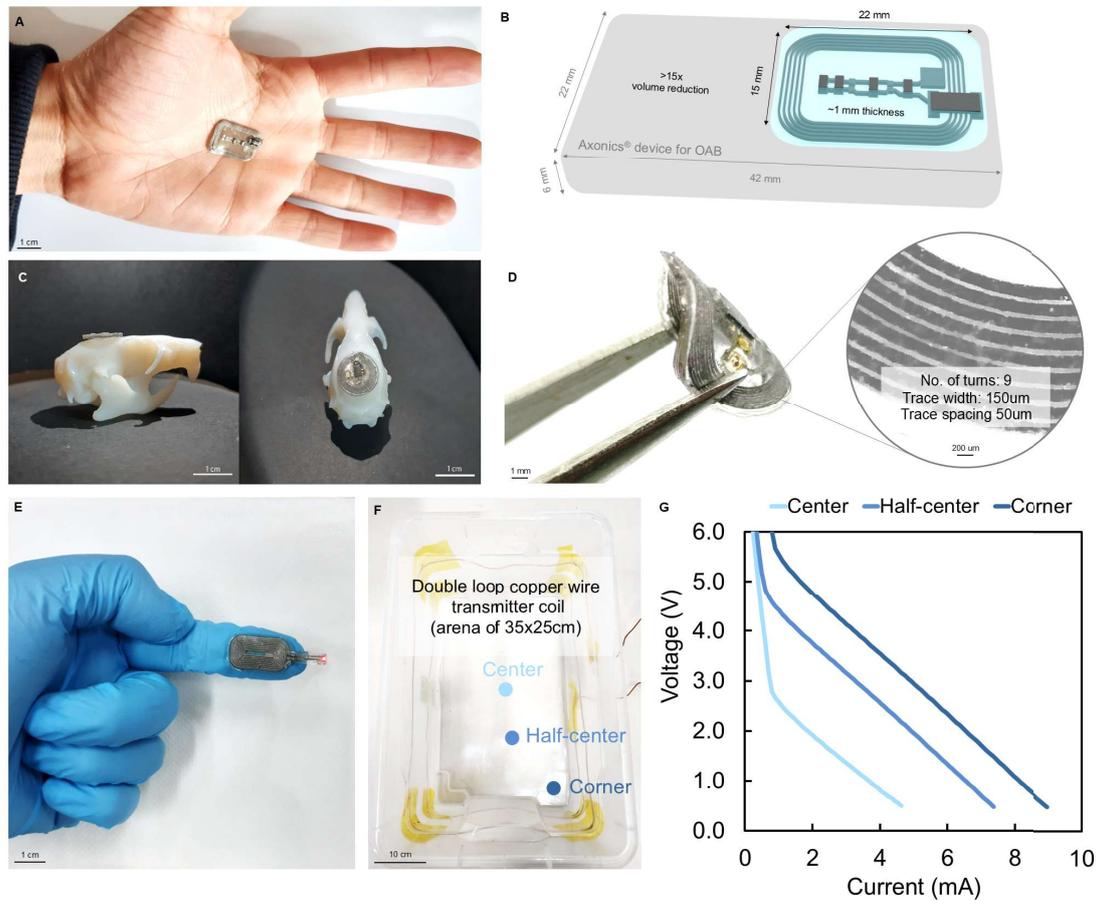

**Figure 5.** Display of the features of the final devices. (A) Photograph of the final device for OAB comparing its size to a hand and a schematic rendering comparing its volume to the one of the commercially available Axonics® device[76]. (C) Size of the smallest developed device in comparison with a 3D printed skull of a rat[77] and in (D) a photograph of its conformability accompanied by a microscope photograph of the fine features of its coil. (E) Photograph of a battery-free wireless device based on liquid metal composite inks optimized for long-range settings (cage described in (F)) for the context of optogenetics studies. (G) Measurements of the electrical performance of the energy harvesting coil of (E) in different locations of the cage (F).